\documentclass[conference]{IEEEtran}
\IEEEoverridecommandlockouts
\usepackage{cite}
\usepackage{amsmath,amssymb,amsfonts}
\usepackage{algorithmic}
\usepackage{graphicx}
\usepackage{textcomp}
\usepackage{xcolor}
\def\BibTeX{{\rm B\kern-.05em{\sc i\kern-.025em b}\kern-.08em
    T\kern-.1667em\lower.7ex\hbox{E}\kern-.125emX}}
\begin{document}

\title{Enhancing Edge Intelligence with Highly Discriminant LNT Features\\
% {\footnotesize \textsuperscript{*}Note: Sub-titles are not captured in Xplore and
% should not be used}
% \thanks{Identify applicable funding agency here. If none, delete this.}
}

\author{\IEEEauthorblockN{Xinyu Wang}
\IEEEauthorblockA{\textit{University of Southern California}\\
Los Angeles, CA, USA \\
xwang350@usc.edu % email address or ORCID
}
\and
\IEEEauthorblockN{Vinod~K.~Mishra}
\IEEEauthorblockA{\textit{DEVCOM Army Research Laboratory}\\
Adelphi, MD, USA \\
vinod.mishra.civ@army.mil}
\and
\IEEEauthorblockN{C.-C. Jay Kuo}
\IEEEauthorblockA{\textit{University of Southern California}\\
Los Angeles, CA, USA \\
jckuo@usc.edu}

% \and
% \IEEEauthorblockN{4\textsuperscript{th} Given Name Surname}
% \IEEEauthorblockA{\textit{dept. name of organization (of Aff.)} \\
% \textit{name of organization (of Aff.)}\\
% City, Country \\
% email address or ORCID}
}

\maketitle

\begin{abstract}

AI algorithms at the edge demand smaller model sizes and lower
computational complexity. To achieve these objectives, we adopt a green
learning (GL) paradigm rather than the deep learning paradigm.  GL has
three modules: 1) unsupervised representation learning, 2) supervised
feature learning, and 3) supervised decision learning. We focus on the
second module in this work. In particular, we derive new discriminant
features from proper linear combinations of input features, denoted by
${\bf x}$, obtained in the first module. They are called complementary
and raw features, respectively.  Along this line, we present a novel
supervised learning method to generate highly discriminant complementary
features based on the least-squares normal transform (LNT). LNT consists
of two steps.  First, we convert a $C$-class classification problem to a
binary classification problem. The two classes are assigned with 0 and
1, respectively.  Next, we formulate a least-squares regression problem
from the $N$-dimensional ($N$-D) feature space to the 1-D output space,
and solve the least-squares normal equation to obtain one $N$-D normal
vector, denoted by ${\bf a}_1$.  Since one normal vector is yielded by
one binary split, we can obtain $M$ normal vectors with $M$ splits.
Then, ${\bf A} {\bf x}$ is called an LNT of ${\bf x}$, where transform
matrix ${\bf A} \in R^{M \times N}$ by stacking ${\bf a}_j^T$, $j=1,
\cdots , M$, and the LNT, ${\bf A} {\bf x}$, can generate $M$ new
features.  The newly generated complementary features are shown to be
more discriminant than the raw features. Experiments show that the
classification performance can be improved by these new features. 

\end{abstract}


\section{Introduction}\label{sec:introduction}

Deep learning (DL) \cite{DL_introduction} has been dominating in the
field of artificial intelligence (AI) and machine learning (ML) in the
last decade. However, its black-box nature and heavy computational
complexity are concerns to academia and industry. In mobile and edge
computing platforms, it is critical to develop lightweight AI/ML models
to enhance edge intelligence.  To build interpretable, reliable, and
lightweigt learning systems, green learning (GL)
\cite{kuo2016understanding, kuo2017cnn, kuo2018data, Saab, DFT, GL} has
been proposed as an alternative to deep learning in recent years. 

GL systems have neither computational neurons nor networks. Instead,
they are composed by three modules: 1) unsupervised representation
learning, 2) supervised feature learning, and 3) supervised decision
learning.  A rich set of representations is generated in the first
module of GL without any supervision. They are rank-ordered according to
their discriminant power by the discriminant feature test (DFT)
\cite{DFT} with supervision in the second module.  Then, discriminant
representations are chosen as features, which are fed to the third
module for final decision-making.  Readers are referred to \cite{GL} for
an extensive overview on this topic. We focus on the second module in
this work. 

Since there is no supervision adopted in Module 1, representations of GL
may not be as competitive as those obtained by end-to-end optimized
DL solutions. To address this shortcoming, an idea of creating new
features by linearly combining selected features was investigated in
\cite{fu2022subspace, fu2022acceleration}. On the one hand, it was shown
that it is feasible get more discriminant features through linear
combinations.  On the other hand, it was a nontrivial problem in finding
the weights of good linear combinations. Several search algorithms were
examined in \cite{fu2022subspace, fu2022acceleration}, including the
probabilistic search, the adaptive particle swarm optimization (APSO)
search, and the stochastic gradient descent (SGD) search. However, they
are still computationally expensive. 

The derived and original input features are called complementary and raw
features, respectively.  A novel and efficient supervised method to
generate discriminant complementary features efficiently is proposed in
this work.  It is named the least-squares normal transform (LNT).  LNT
consists of two steps.  First, a $C$-class classification problem is
converted to a binary classification problem by merging individual
classes into two major super-classes. They are assigned with 0 and 1,
respectively. Next, we formulate a least-squares regression problem from
the $N$-dimensional ($N$-D) feature space to the 1-D output space, and
solve the least-squares normal equation to obtain a $N$-D normal vector,
denoted by ${\bf a}_1$. Since one normal vector is yielded by one binary
split, we have $M$ normal vectors with $M$ splits. Then, ${\bf A} {\bf
x}$ is called an LNT of ${\bf x}$, where transform matrix ${\bf A} \in
R^{M \times N}$ by stacking ${\bf a}_j^T$, $j=1, \cdots, M$, and the
LNT, ${\bf A} {\bf x}$, can generate $M$ new features.  The newly
generated complementary features are shown to be more discriminant than
the raw features.  Experiments show that the classification performance
can be improved by these new features. 

This work has three major contributions. First, a novel LNT method is
proposed as an efficient tool to generate more discriminant
complementary features based on raw features. Second, an SVD-based
low-rank LNT method is presented to lower the computational complexity
and the model size furthermore. Third, the application of LNT to image
classification problems is presented to demonstrate its power in
boosting the classification performance in practice. 

The rest of this paper is organized as follows. The related background
is reviewed in Sec. \ref{sec:Background}. The LNT method and its
application for new complementary feature generation are described in
Sec. \ref{sec:method}.  Experimental results with several classical
image classification problems are shown in Sec.  \ref{sec:experiment}.
Finally, the concluding remarks and future extensions are given in Sec.
\ref{sec:conclusion}. 

\section{Background Review}\label{sec:Background}

\subsection{Edge Intelligence}

Edge intelligence enables the deployment of AI/ML algorithms to edge
devices \cite{zhou2019edge, deng2020edge, xu2021edge}.  A large amount
of data are generated at geographically distributed sensors nowadays.
One example is the Internet of Things (IoT). It is desired to analyze
captured local data, extract the essential information from them, and
transmit summarized information for high-level information fusion. Thus,
edge Intelligence has the potential to provide AI services for an
individual person and/or device. Here, we consider the image
classification problem at edges. 

Neural-network-based image classification methods have evolved for
several decades, including early multi-layer perceptions (MLP)
\cite{rosenblatt1958perceptron, lin2020two}, convolutional neural
networks (e.g., LeNet \cite{lenet5} AlexNet \cite{alexnet}, Resnet
\cite{resnet}, Densenet \cite{huang2017densely}, etc.), and the Vision
Transformer (ViT) \cite{vaswani2017attention,vit}.  To fit deep neural
networks to edge applications, a wide range of pruning and quantization
techniques have been proposed to simplify trained networks while
preserving the performance \cite{liang2021pruning, vadera2022methods}.
Although it is feasible to reduce the number of model parameters and the
computational complexity, it is challenging to shrink them by an order
of magnitude. 

\subsection{Green Learning}

One concern with deep learning is its high carbon footprint. GL has been
developed to address this issue. GL systems are characterized by smaller
model sizes, lower computational complexity in both training and
inference stages (thus, a lower carbon footprint). GL has been applied
to a range of applications with competitive performance.  Successful
GL's application examples include: image classification \cite{ssl,
chen2020pixelhop++, Epixelhop}, texture and image generation
\cite{lei2021tghop, lei2022genhop, azizi2022pager}, low-resolution face
recognition \cite{rouhsedaghat2021low}, face gender classification
\cite{rouhsedaghat2021facehop}, deepfake detection
\cite{chen2021defakehop, chen2022defakehop++}, blind image quality
assessment \cite{mei2022greenbiqa}, anomaly detection
\cite{zhang2021anomalyhop}, image forensics \cite{zhu2022pixelhop,
RGGID, zhu2023green}, graph learning
\cite{xie2022graphhop,xie2023label}, disease classification
\cite{liu2021voxelhop}, point cloud classification, segmentation and
registration \cite{zhang2020pointhop, kadam2022r, zhang2022gsip,
kadam2022greenpco, kadam2023s3i}. GL can be conveniently deployed on
mobile and edge devices. It can also save a significant amount of energy
consumption at cloud centers. 

As mentioned earlier, GL consists of three modules: 1) unsupervised
representation learning, 2) supervised feature learning, and 3)
supervised decision learning.  Although our current work focuses on the
second module, it also involves the first module. These two modules will
be carefully examined below. It is worthwhile to mention that
neural-networks are end-to-end optimization systems. They have no clear
boundary between feature extraction and decision-making. Generally
speaking, earlier and later layers play roles for feature extraction and
decision-making, respectively.  Yet, research was conducted
\cite{xu2017understanding} to shed light on the role of deep features. 

\subsubsection{Unsupervised Representation Learning}

One example of unsupervised representation learning is the histogram of
oriented gradients (HoG) \cite{HOG, yang2022design}. The HoG features
were popular in computer vision before the deep learning era. Their main
limitation is that they do not capture long range dependency
effectively. The unsupervised representation learning of GL is built
upon two ingredients: 1) the Saab transform \cite{Saab} and 2)
successive subspace learning (SSL) \cite{ssl} via the cascade of
channel-wise (C/w) Saab transforms \cite{chen2020pixelhop++} in multiple
stages/scales.  The Saab transform is a variant of principal component
analysis (PCA).  It computes the mean of pixel values of a local patch
and treats it as the DC (direct-current) component. The DC-removed pixel
values of local patches are zero-mean random vectors. Thus, PCA can be
applied, yielding multiple AC (alternating current) components.  They
correspond to the DC and AC Saab coefficients, respectively. The DC and
low-frequency AC coefficients can be down-sampled to a coarser grid,
known as the pooling operation. The max-pooling is used in DL while the
absolute max-pooling is more effective in GL \cite{yang2022design}.  The
Saab transform is a data-driven transform.  It is neither hand-crafted
by humans nor derived from labels through back propagation.  The
multi-stage Saab transforms are conducted in a one-pass feedforward
manner. 

\subsubsection{Supervised Feature Learning}

A rich set of representations is created in the first module.  They are
rank-ordered according to their discriminant power by the discriminant
feature test (DFT) \cite{DFT} with supervision in the second module.  If
the representation set does not contain highly discriminant ones, DFT
cannot generate new features.  One way to derive new discriminant
features is to conduct linear combinations of existing features.  It is
an open question to determine the weights efficiently. In the next
section, we present a supervised method to yield new discriminant
features. 

\section{Proposed LNT Method}\label{sec:method}

The LNT method in the context of image classification is proposed in
this section. First, we describe a pre-processing step that converts a
classification problem to a regression problem in Sec.
\ref{subsec:preprocessing}.  Next, the LNT algorithm is presented in
Sec.  \ref{subsec:LNT}. Finally, to rank the discriminant power of
generated complementary features, a post-processing step is discussed in
Sec.  \ref{subsec:postprocessing}. 

\subsection{Pre-processing: Mapping from Class Labels to Binary 
Values}\label{subsec:preprocessing}

We can build a link between the multi-object classification problem and
the regression problem as follows. For $C$ object classes, unit vectors
in the $R^C$ space define $C$ one-hot vectors, each of which is used to
denote an object class. If each input image has $N$ features, the
classification problem can be formulated as a nonlinear regression
problem in form of $f: R^N \rightarrow R^C$. Although a nonlinear
regression function is needed for higher classification accuracy, we
consider its linear approximation here for two reasons. First, we are
only interested in generating new features here and will leave the
nonlinear operation to the tree-based classifier (e.g., the XGBoost
classifier).  Second, it is desired to have an efficient computational
algorithm. The linear regression problem is well studied and allows fast
computation. 

We can rewrite $f: R^N \rightarrow R^C$ as
\begin{equation}\label{eq:mapping_matrix}
A {\bf x}_j + {\bf b} \approx {\bf w}_j, \quad j=1, \cdots, C,
\end{equation}
where $A \in R^{C \times N}$ is a matrix, ${\bf b} \in R^C$ is a bias
vector, ${\bf x}_j \in R^N$ and ${\bf w}_j \in R^C$ denote a feature
vector of an image input belonging to the $j$th class and the one-hot
vector associated with the $j$th class, respectively. Furthermore, if 
${\bf a}_i^T$ is the $i$th row of matrix $A$, we have
\begin{equation}\label{eq:mapping_vector}
{\bf a}_i^T {\bf x}_j + {\bf b}_i \approx \delta_{i,j}, \quad 
i,j=1, \cdots, C,
\end{equation}
where $\delta_{i,j}$ is the Kronecker delta. It is equal to one if
$i=j$. Otherwise, it is equal to zero.  Clearly, ${\bf a}_i^T \in R^N$
is the weight vector of a linear combination of input features.  The
Kronecker delta in Eq. (\ref{eq:mapping_vector}) can be interpreted as a
one-versus-others split. If the number, $L$, of training samples is
greater than $N$, ${\bf a}_i$ can be solved using the linear
least-squares regression as presented in Sec. \ref{subsec:LNT}. 

To enrich the set of weight vectors, we adopt various grouping schemes,
such as ``1 vs. (C-1)", ``2 vs. (C-2)", $\cdots$, ``k vs. (C-k)", where
$k=\lfloor C/2 \rfloor $.  For a specific ``l vs. (C-l)" grouping, we
have ${C \choose l}$ splits. As a result, the total number of splits can
be computed via
% \begin{equation}\label{eq:split}
% M = \sum_{l=1}^k {C \choose l}, \qquad k=\lfloor C/2 \rfloor.
% \end{equation}
\begin{equation}\label{eq:split}
M=\left\{
\begin{array}{lcl}
\sum_{l=1}^k {C \choose l}     &      & \text{C is odd}, k=\lfloor C/2 \rfloor.\\
\\
\sum_{l=1}^k {C \choose l} -  \frac{1}{2}\times{C \choose k}  &      & \text{C is even}, k=C/2.
\end{array}
\right.
% M = \sum_{l=1}^k {C \choose l} - \frac{1}{2}\times{C \choose 2}, \qquad k=\lfloor C/2 \rfloor.
\end{equation}
To give an example, we have $C=10$ and $M=512$ for several commonly used
datasets such as MNIST and CIFAR-10. One split partitions $C$ classes
into two super-classes with labels ``0" and ``1".

%%%%%%%%%%%%%%%%%%%%%%%%%%%%%%%%%%%%%%%%%%%%%%%%%%%%%%%
\begin{figure*}[htbp]
\center{
\includegraphics[width=0.3\textwidth,height=4cm]{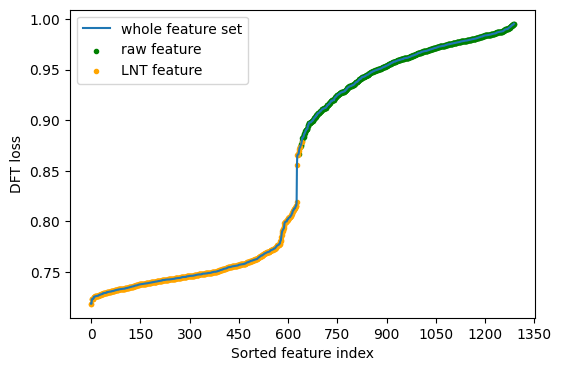}
\includegraphics[width=0.3\textwidth,height=4cm]{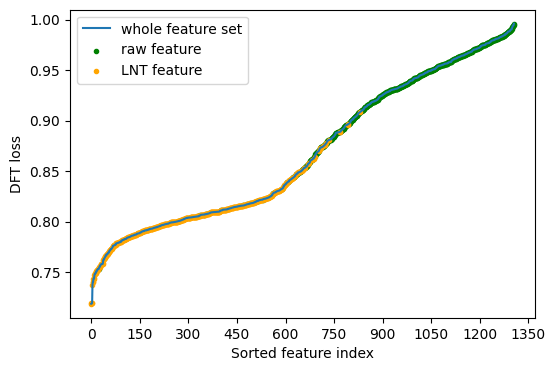}
\includegraphics[width=0.3\textwidth,height=4cm]{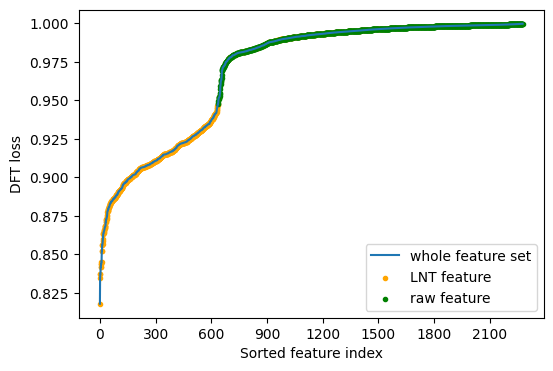}}
\caption{Comparison of the discriminant power of raw and generated features
of three datasets (from left to right): MNIST, Fashion MNIST, and CIFAR-10.
The DFT loss values of raw and generated features are shown in green and orange 
points in the curve, respectively. A feature is more discriminant if its DFT loss
value is smaller.}\label{fig:DFT}
\end{figure*}
%%%%%%%%%%%%%%%%%%%%%%%%%%%%%%%%%%%%%%%%%%%%%%%%%%%%%%%

%Jay: please replace the above three figures with DFT loss figures.
%     Name them mnist_dft.png, fashion_dft.png, and cifar10_dft.png

%%%%%%%%%%%%%%%%%%%%%%%%%%%%%%%%%%%%%%%%%%%%%%%%%%%%%%%
\begin{figure*}[htbp]
\centerline{\includegraphics[width=0.9\textwidth]{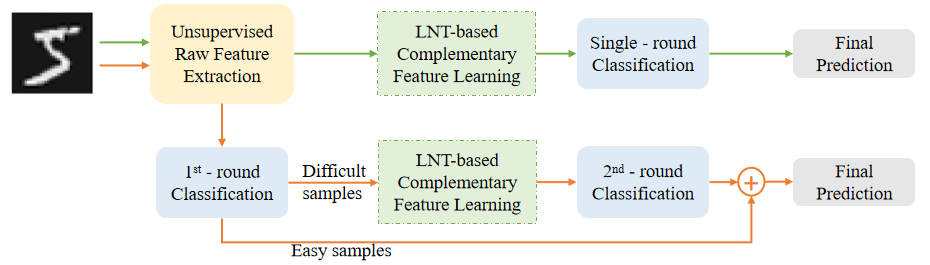}}
\caption{The block diagram of the image classification system, which
follows the standard GL data processing pipeline. We propose one-round
of two-round decision systems and indicate them in green and red arrows,
respectively.}\label{fig:system}
\end{figure*}
%%%%%%%%%%%%%%%%%%%%%%%%%%%%%%%%%%%%%%%%%%%%%%%%%%%%%%%

\subsection{Least-Squares Normal Transform (LNT)}\label{subsec:LNT}

We associate each split with a target vector ${\bf t} \in R^C$, whose
elements are either zero or one. They serve as indicators for one of the
two super classes. With $L$ training samples, we can formulate the
following linear least-squares regression problem to solve for $M$
weight vectors simultaneously:
\begin{equation}\label{eq:lls}
AX + B = T, 
\end{equation}
where $A \in R^{M \times N}$ is the weight matrix, $X \in R^{N \times
L}$ is the data matrix, $B \in R^{M \times L}$ is the bias matrix, and
$T \in R^{M \times L}$ is the indicator matrix. Since all training
samples share the same bias vector, $B$ is a rank-one matrix with
the same column vector repeated $L$ times. The element $t_{m,l}$ in
$T$ takes values 0 or 1 depending on split scheme $1 \le m \le M$ and
the super-class label of the $l$th training sample. 

By taking the expectation of Eq. (\ref{eq:lls}), we have
\begin{equation}\label{eq:bias}
B = E[T] - A E[X].
\end{equation}
Actually, we are not concerned with $B$ but $A$ since we are only interested
in generating new complementary features. Since the bias term shifts a
feature with a certain value, it has no effect on its discriminant
power. Here, we focus on the solution of weight matrix $A$ only. 
Matrix $A$ can be obtained by solving the least-squares normal equation.
It has a closed form solution in form of
\begin{equation}\label{eq:lneq}
A = TX^T (X X^T)^{-1}.
\end{equation}

For a training or testing sample with its raw feature vector ${\bf x}$,
its $M$ complementary features can be generated via
\begin{equation}\label{eq:lnt}
A {\bf x} = {\bf p},
\end{equation}
where matrix $A$ is obtained from Eq. (\ref{eq:lneq}) and ${\bf p}=(p_1,
\cdots, p_M)$, and where $p_m$ is the value of the $m$th generated
feature. Eq. (\ref{eq:lnt}) is called the least-squares normal transform
(LNT). 

\subsection{Post-processing: Selection of Discriminant Generated Features}
\label{subsec:postprocessing}

After feature generation, we apply the discriminant feature test (DFT)
\cite{DFT} to complementary features to find their discriminant power.
DFT is a semi-supervised feature selection method. It partitions the
dynamic range of a 1-D feature into two non-overlapping intervals and
computes the weighted entropy loss from them. Furthermore, it searches
for the minimum weighted entropy loss among a set of uniformly sampled
points and uses it as the DFT loss of the feature. A feature is more
discriminant if its DFT loss value is smaller. One sort all features
according to their DFT loss values from the smallest to the largest to
obtain a DFT loss curve (see Fig. \ref{fig:DFT}) and use the elbow point
of the DFT curve to select a set of discriminant features for decision
making. We compare the discriminant power of raw and newly generated
complementary features for the MNIST, Fashion-MNIST, and CIFAR-10 three
datasets in Fig. \ref{fig:DFT}. The DFT loss values of raw and generated
features are shown in green and orange points in the curve, respectively.
We see clearly that the generated complementary features have higher
discriminability than the raw features. 

\subsection{SVD-based Low-Rank LNT Method}\label{subsection:SVD-Enhanced}

Addressing the least-squares regression problem through LNT yields
weight matrix $A \in R^{M \times N}$, where $M$ denotes the number of
super-classes considered in the LNT framework. Each row of matrix $A$ is
an individual LNT filter responsible for the generation of a distinct
complementary feature. Consequently, each super-class corresponding to a
unique LNT filter yields a specific complementary feature. Under this
formulation, the dimension reduction of complementary features results
in a commensurate exclusion of $K$ super-classes. Even we perform
discriminant feature selection via DFT, the reduction in feature
dimensions still significantly affects the final performance. To
maintain the performance while achieving a reasonable feature dimension,
we propose an enhanced LNT method that enables each filter to consider
multiple super-classes, ensuring that feature dimension reduction does
not significantly impact the performance. 

After obtaining weight matrix $A \in R^{M \times N}$ via LNT by following
the steps described in Sections \ref{subsec:preprocessing} and
\ref{subsec:LNT}, we apply the singular value decomposition (SVD) to
$A^T$:
\begin{equation}\label{eq:svd}
A^T = U \Sigma V^T,
\end{equation}
where $U \in R^{N \times N}$ and $V \in R^{M \times M}$ are unitary
matrices containing the left-singular and right-singular vectors, and
$\Sigma \in R^{N \times M}$ contains singular values along the diagonal.
The importance of each singular vector is ordered by its corresponding
singular value. The dimension of weight matrix $A$ can be reduced by
discarding singular vectors associated with the smallest eigenvalues. 

Matrices $U$, $\Sigma$, and $V$ are then truncated by retaining only the
first k columns (or rows) in each matrix. That is, we obtain
dimension-reduced matrices $U_k \in R^{N \times K}$, $\Sigma_k \in R^{K
\times K}$, and $V_k \in R^{M \times K}$. Then, the low-rank
approximation of $A$ can be written as
\begin{equation}\label{eq:svd2}
\tilde{A}^T_k = U_k \Sigma_k V_k^T.
\end{equation}
Matrix $\tilde{A}^T_k$ preserves the most significant components of $A$
and serves as the weight matrix in a lower-dimensional space.  Each
remaining singular vector can be constructed by taking all rows of $A$
into account, encompassing the information of all $M$ super-classes.
The SVD-based low-rank LNT approximation method minimizes the impact of
dimension reduction on the classification performance. 

It is worthwhile to comment that original LNT filters span the row space
of $A$ and the above procedure considers dimension reduction in the row
space. As an optional step, further simplification of the low-rank
approximation $\tilde{A}_k$ can be achieved by leveraging its column
space, leading to matrix $\tilde{A'}_k'$.  Finally, each row of the
dimension-reduced matrix $\tilde{A'}_k'$ can be employed as the newly
generated LNT filters. 

\section{Experiments}\label{sec:experiment}

We demonstrate the performance improvement of image classification systems
by using the proposed LNT method in this section. 

\subsection{Experimental Setup}

We conduct experiments on MNIST \cite{MNIST}, Fashion-Mnist (F-Mnist)
\cite{xiao2017fashion}, and CIFAR-10\cite{cifar10} three datasets. 
The image classification system is shown in Fig. \ref{fig:system}. It
follows the standard GL data processing pipeline. A 3-hop pixelhop++
\cite{chen2020pixelhop++} architecture is used for unsupervised raw
feature extraction. Its hyper-parameters are given in Table
\ref{tab1:hyper_pixelhop}.  Only the Saab coefficients in the last hop
are used as raw features. For color images, we covert the RGB channels
to YUV channels, apply the same 3-hop pixelhop++ to each channel, and
use all of them as raw features.  Furthermore, the HOG features are
included in the raw feature set for F-Mnist and CIFAR-10 to increase
feature diversity. 

%%%%%%%%%%%%%%%%%%%%%%%%%%%%%%%%%%%%%%%%%%%%%%%%%%%
\begin{figure*}[htbp]
\center
{\includegraphics[width=0.3\textwidth,height=5cm]{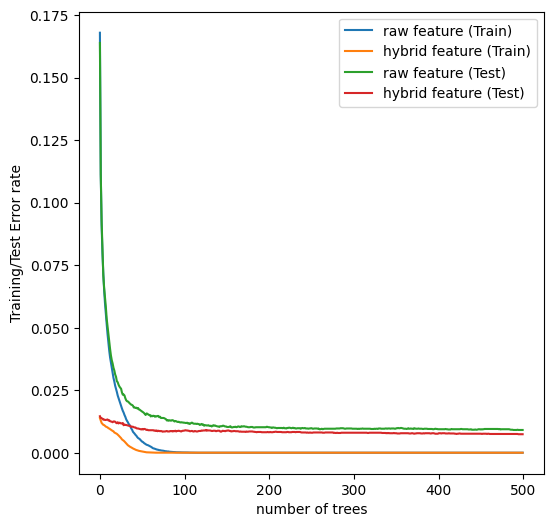}}
{\includegraphics[width=0.3\textwidth,height=5cm]{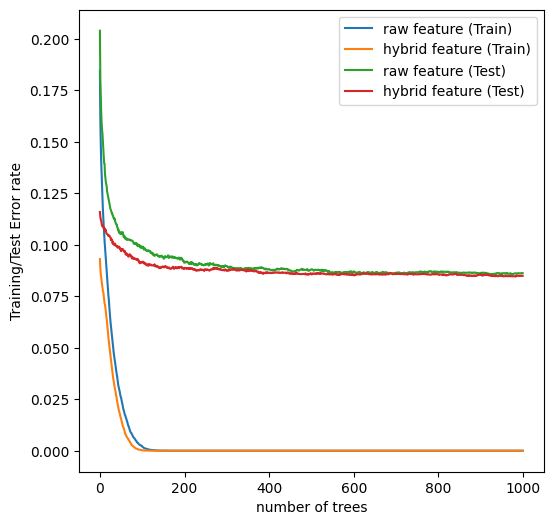}}
{\includegraphics[width=0.3\textwidth,height=5cm]{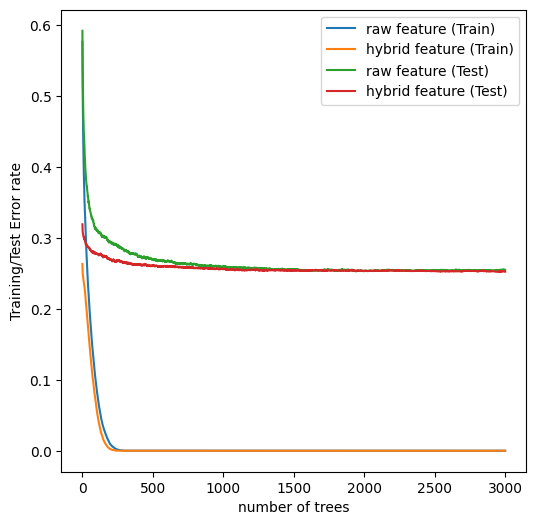}}
\caption{The error rate as a function of the tree number in baseline and
one-round system for MNIST (left), F-Mnist (middle) and CIFAR-10 (right). 
For comparison, we set the same tree depth for the baseline and the 
one-round system with basic LNT.} \label{fig: converge}
\end{figure*}
%%%%%%%%%%%%%%%%%%%%%%%%%%%%%%%%%%%%%%%%%%%%%%%%%%%

% %%%%%%%%%%%%%%%%%%%%%%%%%%%%%%%%
\begin{table}[htbp]
\caption{The hyper-parameters of the pixelhop++ architecture for 
MNIST, F-Mnist and CIFAR-10.}\label{tab1:hyper_pixelhop}
\begin{center}
\begin{tabular}{c|c|c|c} \hline
 & \multicolumn{3}{c} {Filter Spatial Size (Channel \#)}\\ \cline{2-4}
  & MNIST & F-Mnist & CIFAR-10\\ \hline
  % & {Spatial Size} & {Spatial Size} & {Spatial Size}\\
  % & (Channel \#) & (Channel \#) & (Channel \#)\\ \hline
 Hop-1 &  5x5 (25) & 5x5 (25)  & 5x5 (69) \\
 Hop-2 & 5x5 (256) & 5x5 (210)  & 5x5 (576)  \\
 Hop-3& 5x5 (652) & 5x5 (469) & 5x5 (1441) \\ \hline
 % Total para \# &  23.325k &  17.6k &  k \\ 
\end{tabular}
\end{center}
\end{table}
% %%%%%%%%%%%%%%%%%%%%%%%%%%%%%%%%

In the one-round system, we apply LNT to the original feature set
directly to generate complementary features, and train an XGBoost
classier with DFT selected features to predict the image class. In the
two-round system, we use the raw feature set for decision making in the
first round. Then, we categorize samples into easy and difficult ones
based on their confidence scores. The confidence score is calculated as
the entropy of the predicted label vector. A higher entropy implies a
less confident decision. The round-one decision of an easy test sample
is accepted as is. On the other hand, a difficult test sample will be
fed into the second-round decision pipeline, where LNT is used to
generate complementary features and another XGBoost classifier is
trained. The prediction in the second round is accepted as the
prediction of difficult samples. 

In the one-round system, all samples go through the LNT process so that
its computational complexity in the inference stage is higher. In the
two-round system, we can lower the inference complexity since easy
samples do not go through the LNT process.  On the other hand, the
two-round system has a larger model size (i.e., more model parameters)
than the one-round system. Thus, the one- and two-round designs provide
a trade-off between computational and storage complexities. Furthermore,
for the performance benchmarking purpose, the one-round system with raw
features only (i.e., without LNT generated features) is called the
baseline method. 

%%%%%%%%%%%%%%%%%%%%%%%%%%%%%%%%
\begin{table}[htbp]
\caption{Hyper parameters and model sizes of the XGBoost classifiers.}
\label{tab:xgboost}
\begin{center}
\begin{tabular}{c|c|c|c|c} \hline
 % & \multicolumn{2}{c||} {Fashion Mnist} & \multicolumn{2}{|c}{MNIST} \\ \hline
Dataset & Setting & LNT type & depth(tree \#) & Model Size \\ \hline
MNIST & baseline & & 3(500) & 110k  \\
MNIST & 1-round & Basic & 2(500)  & 50k  \\
MNIST & 2-round & Basic & 2(500+200) & 70k  \\
MNIST & 1-round & Low-rank & 2(500) & 50k  \\
MNIST & 2-round & Low-rank & 2(350+200) & 55k  \\ \hline
F-Mnist & baseline & & 5(1000) & 940k \\ 
F-Mnist & 1-round & Basic & 3(2000) & 440k \\\
F-Mnist & 2-round & Basic & 4(500+1200) & 782k \\
F-Mnist & 1-round & Low-rank & 3(1500) & 330k \\\
F-Mnist & 2-round & Low-rank & 4(500)+3(2000) & 670k \\ \hline
CIFAR-10 & baseline & & 4(3000) & 1380k\\
CIFAR-10 & 1-round & Basic & 3(3000) & 660k\\
CIFAR-10 & 2-round & Basic & 4(1500+1500) & 1380k\\
CIFAR-10 & 1-round & Low-rank & 3(2500)  & 550k\\
CIFAR-10 & 2-round & Low-rank & 4(1500)+3(2000) & 1130k\\ \hline
\end{tabular}
\end{center}
\end{table}
%%%%%%%%%%%%%%%%%%%%%%%%%%%%%%%%

\subsection{Performance Evaluation}

We compare the classification performance of six methods; namely,
LeNet-5 \cite{lenet5}, PixelHop \cite{ssl}, PixelHop$^+ $\cite{ssl}, the
baseline, one-round and two-round classification systems. In the
one-round system, all training samples are used to obtain the LNT
features. In the two-round system, we select $c_1$\% training samples
and $c_2$\% testing samples to go to the second round. The experiments
employ either the basic LNT method or the SVD-based low-rank LNT method
in the designed systems. Both methods includes preprocessing, LNT, and
post-processing steps as outlined in sections
\ref{subsec:preprocessing}, \ref{subsec:LNT} and
\ref{subsec:postprocessing}, while an additional dimension reduction
step described in \ref{subsection:SVD-Enhanced} is adopted in the
SVD-based low-rank LNT method. 

For the basic LNT method, we choose $c_1=20, \, 60, \, 80$ and $c_2 = 2,
\, 20, \, 20$ for MNIST, F-Mnist, CIFAR-10, respectively. Additionally,
we choose 200 most discriminant complementary features for MNIST and use
all 512 complementary features for F-Mnist and CIFAR-10. In contrast,
the low-rank LNT method employs different $c1$ and $c2$ values, where
$c_1=20, \, 60, \, 80$ and $c_2 = 2.5, \, 30, \, 20$ for MNIST, F-Mnist,
CIFAR-10, respectively. Furthermore, we only generate 200 complementary
features for each dataset.  Subsequently, we take the union of raw and
complementary features and feed them into the XGBoost classifier in the
final stage. The hyper parameters of the XGBoost classifiers used in the
baseline, one-round and two-round systems are summarized in Table
\ref{tab:xgboost}. 

%%%%%%%%%%%%%%%%%%%%%%%%%%%%%%%%
\begin{table}[htbp]
\caption{Comparison of classification accuracy on MNIST, F-Mnist, CIFAR-10 with
LeNeT-5, PixelHop, PixelHop$^+$, the proposed one-round and two-round systems.}
\label{tab3: performance}
\begin{center}
\begin{tabular}{c|c|c|c}
\hline
 % & \multicolumn{2}{c||} {Fashion Mnist} & \multicolumn{2}{|c}{Mnist} \\ \hline
 & \multicolumn{3}{c} {Test Accuracy (\%)}\\ 
\cline{2-4}
  & MNIST & F-Mnist & CIFAR-10\\ \hline
LeNet-5 & 99.07 & 89.54 & 68.72 \\
PixelHop & 98.90 & 91.30 & 71.37\\
PixelHop$^+$ & 99.09 & 91.68 & 72.66\\ \hline
Baseline (Ours) & 99.09 & 91.38 & 74.51 \\ 
1-round (Ours - Basic LNT) & 99.24 & 91.75 & 74.80 \\ 
2-round (Ours - Basic LNT) & \textbf{99.32} & \textbf{92.07} & \textbf{75.93}\\ \hline
1-round (Ours - Low-rank LNT) & 99.28 & 91.72 & 75.68\\ 
2-round (Ours - Low-rank LNT) & \textbf{99.33} & \textbf{92.03} & \textbf{76.28}\\ \hline
\end{tabular}
\end{center}
\end{table}
%%%%%%%%%%%%%%%%%%%%%%%%%%%%%%%%

%%%%%%%%%%%%%%%%%%%%%%%%%%%%%%%%
\begin{table}[htbp]
\caption{Comparison of model sizes in the complementary feature
generation process with basic LNT and low-rank LNT methods.} 
\label{tab:model_size}
\begin{center}
\begin{tabular}{c|c|c|c|c}\hline
& LNT type & MNIST & F-Mnist & CIFAR-10\\ \hline
1-round system & Basic & 102.4k & 240.1k & 737.8k \\ 
2-round system & Basic & 102.4k & 240.1k& 737.8k \\ \hline
1-round system & Low-rank & 102.4k & 93.8k & 288.2k\\ 
2-round system & Low-rank & 102.4k & 93.8k & 288.2k \\ \hline
\end{tabular}
\end{center}
\end{table}
%%%%%%%%%%%%%%%%%%%%%%%%%%%%%%%%

%%%%%%%%%%%%%%%%%%%%%%%%%%%%%%%%
\begin{table}[htbp]
\caption{Comparison of computational complexity in the inference stage
in terms of the averaged floating point operations (FLOPs) required to
classify an image, where only computations in the complementary feature
generation and the XGBoost classification are considered.} 
\label{tab:complexity}
\begin{center}
\begin{tabular}{c|c|c|c}\hline
& \multicolumn{3}{c} {FLOPs \#}\\ \cline{2-4}
& MNIST & F-Mnist & CIFAR-10\\ \hline
Baseline system & 20k & 60k & 150k\\
1-round system & 15k+204.8k & 80k+480.2k & 120k+1475.6k \\ 
2-round system & 21k+4.1k & 85k+144.06k & 150k+295.12 \\ \hline
1-round system & 15k+204.8k & 60k+187.6k & 100k+576.4k\\ 
2-round system & 16.5k+5.12k & 105k+56.28k & 155k+115.28k \\ \hline
\end{tabular}
\end{center}
\end{table}
%%%%%%%%%%%%%%%%%%%%%%%%%%%%%%%%

We conduct performance benchmarking in three measures: 1) classification
accuracy, 2) the number of model parameters (i.e., the model size), and
3) the computational complexity in inference. The results are shown in
Tables \ref{tab3: performance}, \ref{tab:model_size}, and
\ref{tab:complexity}, respectively. We have the following observations. 

\subsubsection{Classification Accuracy} 

We see from Table \ref{tab3: performance} that the use of LNT-generated
complementary features boost the baseline performance to a higher level.
The 2-round and one-round systems with different LNT methods outperforms
the baseline in all datasets. Compared to basic LNT, low-rank LNT can
achieve the same or even better performance.  The 2-round system with
low-rank LNT outperforms the baseline by 0.24\%, 0.65\% and 1.77\% for
MNIST, F-Mnist, and CIFAR-10, respectively.  Furthermore, complementary
features help the XGBoost classifier to converge faster as shown in Fig.
\ref{fig: converge}, where the green and red curves show the convergence
rate in the inference stage of the baseline and the one-round system,
respectively. Clearly, the one-round system converges faster than the
baseline in all datasets. 

\subsubsection{Model Sizes} 

We compare the model sizes of the XGBoost classifiers of the baseline,
one-round and two-round systems in Table \ref{tab:xgboost}. The model
size of the one-round system is around 50\% of that of the baseline.
Although there are two XGBoost classifiers in the two-round system, its
total model size is still slightly lower than that of the baseline. The
advantages of introducing complementary features are demonstrated by
higher classification accuracy and smaller model sizes. As compared with
basic LNT, low-rank LNT helps reduce the number of parameters used in
XGBoost classifiers and reduce the model size by 60\% in F-Mnist and
CIFAR-10 as shown in Table \ref{tab:model_size}. 

\subsubsection{Computational Complexity in Inference} 

We compare the computational complexity in the inference stage in terms
of the averaged floating point operations (FLOPs) required to classify
one image in Table \ref{tab:complexity}. Since these systems share the
same raw feature generation module, we only compare the number of FLOPs
required by complementary feature generation and the XGBoost
classification. Since each image has $32\times32=1,024$ pixels, the
complexity can be normalized per pixel by dividing the numbers in Table
\ref{tab:complexity} with 1024. We see that the two-round system has
much lower FLOPs than the one-round system. It is around 10\%-15\%,
10\%-65\%, 25\%-40\% of the one-round system for MNIST, F-Mnist,
CIFAR-10, respectively. 

\section{Conclusion and Future Work}\label{sec:conclusion}

Based on the least-squares normal transform (LNT), a novel supervised
feature generation method was proposed in this work.  The newly
generated complementary features are more discriminant than raw
features.  Furthermore, an SVD-based low-rank LNT method was proposed to
reduce the number of model parameters and boost the performance.  The
use of LNT features in the image classification task was demonstrated.
It was shown experimentally that the LNT features boosted the
classification performance and improved the convergence behavior of the
XGBoost classifier.  As future extensions, we would like to shed more
light on LNT generated features and apply them in a wider range of image
processing and computer vision tasks. 

\section*{Acknowledgment}

The U.S. Government is authorized to reproduce and distribute reprints
for Governmental purposes notwithstanding any copyright notation
thereon.  The views and conclusions contained herein are those of the
authors and should not be interpreted as necessarily representing the
official policies or endorsements, either expressed or implied, of US
Army Research Laboratory (ARL) or the U.S. Government. Computation for
the work was supported by the University of Southern California's Center
for Advanced Research Computing (carc.usc.edu). 

% \section*{References}
\bibliographystyle{IEEEtran}
\bibliography{ref}

% Generated by IEEEtran.bst, version: 1.14 (2015/08/26)
\begin{thebibliography}{10}
\providecommand{\url}[1]{#1}
\csname url@samestyle\endcsname
\providecommand{\newblock}{\relax}
\providecommand{\bibinfo}[2]{#2}
\providecommand{\BIBentrySTDinterwordspacing}{\spaceskip=0pt\relax}
\providecommand{\BIBentryALTinterwordstretchfactor}{4}
\providecommand{\BIBentryALTinterwordspacing}{\spaceskip=\fontdimen2\font plus
\BIBentryALTinterwordstretchfactor\fontdimen3\font minus \fontdimen4\font\relax}
\providecommand{\BIBforeignlanguage}[2]{{%
\expandafter\ifx\csname l@#1\endcsname\relax
\typeout{** WARNING: IEEEtran.bst: No hyphenation pattern has been}%
\typeout{** loaded for the language `#1'. Using the pattern for}%
\typeout{** the default language instead.}%
\else
\language=\csname l@#1\endcsname
\fi
#2}}
\providecommand{\BIBdecl}{\relax}
\BIBdecl

\bibitem{DL_introduction}
Y.~LeCun, Y.~Bengio, and G.~Hinton, ``Deep learning,'' \emph{Nature}, vol. 521, pp. 436--44, 05 2015.

\bibitem{kuo2016understanding}
C.-C.~J. Kuo, ``Understanding convolutional neural networks with a mathematical model,'' \emph{Journal of Visual Communication and Image Representation}, vol.~41, pp. 406--413, 2016.

\bibitem{kuo2017cnn}
------, ``The {CNN} as a guided multilayer {RECOS} transform,'' \emph{IEEE signal processing magazine}, vol.~34, no.~3, pp. 81--89, 2017.

\bibitem{kuo2018data}
C.-C.~J. Kuo and Y.~Chen, ``On data-driven {Saak} transform,'' \emph{Journal of Visual Communication and Image Representation}, vol.~50, pp. 237--246, 2018.

\bibitem{Saab}
C.-C.~J. Kuo, M.~Zhang, S.~Li, J.~Duan, and Y.~Chen, ``Interpretable convolutional neural networks via feedforward design,'' \emph{Journal of Visual Communication and Image Representation}, vol.~60, pp. 346--359, 2019.

\bibitem{DFT}
Y.~Yang, W.~Wang, H.~Fu, and C.-C.~J. Kuo, ``On supervised feature selection from high dimensional feature spaces,'' \emph{arXiv preprint arXiv:2203.11924}, 2022.

\bibitem{GL}
C.-C.~J. Kuo and A.~M. Madni, ``Green learning: Introduction, examples and outlook,'' \emph{Journal of Visual Communication and Image Representation}, vol.~90, p. 103685, 2023.

\bibitem{fu2022subspace}
H.~Fu, Y.~Yang, V.~K. Mishra, and C.-C.~J. Kuo, ``Subspace learning machine (slm): Methodology and performance,'' in \emph{2023 IEEE International Conference on Acoustics, Speech and Signal Processing (ICASSP)}.\hskip 1em plus 0.5em minus 0.4em\relax IEEE, 2023, pp. 1--5.

\bibitem{fu2022acceleration}
H.~Fu, Y.~Yang, Y.~Liu, J.~Lin, E.~Harrison, V.~K. Mishra, and C.-C.~J. Kuo, ``Acceleration of subspace learning machine via particle swarm optimization and parallel processing,'' in \emph{2022 Asia-Pacific Signal and Information Processing Association Annual Summit and Conference (APSIPA ASC)}.\hskip 1em plus 0.5em minus 0.4em\relax IEEE, 2022, pp. 1019--1024.

\bibitem{zhou2019edge}
Z.~Zhou, X.~Chen, E.~Li, L.~Zeng, K.~Luo, and J.~Zhang, ``Edge intelligence: Paving the last mile of artificial intelligence with edge computing,'' \emph{Proceedings of the IEEE}, vol. 107, no.~8, pp. 1738--1762, 2019.

\bibitem{deng2020edge}
S.~Deng, H.~Zhao, W.~Fang, J.~Yin, S.~Dustdar, and A.~Y. Zomaya, ``Edge intelligence: The confluence of edge computing and artificial intelligence,'' \emph{IEEE Internet of Things Journal}, vol.~7, no.~8, pp. 7457--7469, 2020.

\bibitem{xu2021edge}
D.~Xu, T.~Li, Y.~Li, X.~Su, S.~Tarkoma, T.~Jiang, J.~Crowcroft, and P.~Hui, ``Edge intelligence: Empowering intelligence to the edge of network,'' \emph{Proceedings of the IEEE}, vol. 109, no.~11, pp. 1778--1837, 2021.

\bibitem{rosenblatt1958perceptron}
F.~Rosenblatt, ``The perceptron: a probabilistic model for information storage and organization in the brain.'' \emph{Psychological review}, vol.~65, no.~6, p. 386, 1958.

\bibitem{lin2020two}
R.~Lin, Z.~Zhou, S.~You, R.~Rao, and C.-C.~J. Kuo, ``From two-class linear discriminant analysis to interpretable multilayer perceptron design,'' \emph{arXiv preprint arXiv:2009.04442}, 2020.

\bibitem{lenet5}
Y.~LeCun, B.~Boser, J.~S. Denker, D.~Henderson, R.~E. Howard, W.~Hubbard, and L.~D. Jackel, ``Backpropagation applied to handwritten zip code recognition,'' \emph{Neural computation}, vol.~1, no.~4, pp. 541--551, 1989.

\bibitem{alexnet}
A.~Krizhevsky, I.~Sutskever, and G.~E. Hinton, ``Imagenet classification with deep convolutional neural networks,'' \emph{Advances in neural information processing systems}, vol.~25, 2012.

\bibitem{resnet}
K.~He, X.~Zhang, S.~Ren, and J.~Sun, ``Deep residual learning for image recognition,'' in \emph{Proceedings of the IEEE conference on computer vision and pattern recognition}, 2016, pp. 770--778.

\bibitem{huang2017densely}
G.~Huang, Z.~Liu, L.~Van Der~Maaten, and K.~Q. Weinberger, ``Densely connected convolutional networks,'' in \emph{Proceedings of the IEEE conference on computer vision and pattern recognition}, 2017, pp. 4700--4708.

\bibitem{vaswani2017attention}
A.~Vaswani, N.~Shazeer, N.~Parmar, J.~Uszkoreit, L.~Jones, A.~N. Gomez, L.~Kaiser, and I.~Polosukhin, ``Attention is all you need,'' 2017.

\bibitem{vit}
A.~Dosovitskiy, L.~Beyer, A.~Kolesnikov, D.~Weissenborn, X.~Zhai, T.~Unterthiner, M.~Dehghani, M.~Minderer, G.~Heigold, S.~Gelly \emph{et~al.}, ``An image is worth 16x16 words: Transformers for image recognition at scale,'' \emph{arXiv preprint arXiv:2010.11929}, 2020.

\bibitem{liang2021pruning}
T.~Liang, J.~Glossner, L.~Wang, S.~Shi, and X.~Zhang, ``Pruning and quantization for deep neural network acceleration: A survey,'' \emph{Neurocomputing}, vol. 461, pp. 370--403, 2021.

\bibitem{vadera2022methods}
S.~Vadera and S.~Ameen, ``Methods for pruning deep neural networks,'' \emph{IEEE Access}, vol.~10, pp. 63\,280--63\,300, 2022.

\bibitem{ssl}
Y.~Chen and C.-C.~J. Kuo, ``Pixelhop: A successive subspace learning (ssl) method for object recognition,'' \emph{Journal of Visual Communication and Image Representation}, vol.~70, p. 102749, 2020.

\bibitem{chen2020pixelhop++}
Y.~Chen, M.~Rouhsedaghat, S.~You, R.~Rao, and C.-C.~J. Kuo, ``Pixelhop++: A small successive-subspace-learning-based (ssl-based) model for image classification,'' in \emph{2020 IEEE International Conference on Image Processing (ICIP)}.\hskip 1em plus 0.5em minus 0.4em\relax IEEE, 2020, pp. 3294--3298.

\bibitem{Epixelhop}
Y.~Yang, V.~Magoulianitis, and C.-C.~J. Kuo, ``E-pixelhop: An enhanced pixelhop method for object classification,'' in \emph{2021 Asia-Pacific Signal and Information Processing Association Annual Summit and Conference (APSIPA ASC)}, 2021, pp. 1475--1482.

\bibitem{lei2021tghop}
X.~Lei, G.~Zhao, K.~Zhang, and C.-C.~J. Kuo, ``Tghop: an explainable, efficient, and lightweight method for texture generation,'' \emph{APSIPA Transactions on Signal and Information Processing}, vol.~10, p. e17, 2021.

\bibitem{lei2022genhop}
X.~Lei, W.~Wang, and C.-C.~J. Kuo, ``Genhop: An image generation method based on successive subspace learning,'' in \emph{2022 IEEE International Symposium on Circuits and Systems (ISCAS)}.\hskip 1em plus 0.5em minus 0.4em\relax IEEE, 2022, pp. 3314--3318.

\bibitem{azizi2022pager}
Z.~Azizi, C.-C.~J. Kuo \emph{et~al.}, ``Pager: Progressive attribute-guided extendable robust image generation,'' \emph{APSIPA Transactions on Signal and Information Processing}, vol.~11, no.~1, 2022.

\bibitem{rouhsedaghat2021low}
M.~Rouhsedaghat, Y.~Wang, S.~Hu, S.~You, and C.-C.~J. Kuo, ``Low-resolution face recognition in resource-constrained environments,'' \emph{Pattern Recognition Letters}, vol. 149, pp. 193--199, 2021.

\bibitem{rouhsedaghat2021facehop}
M.~Rouhsedaghat, Y.~Wang, X.~Ge, S.~Hu, S.~You, and C.-C.~J. Kuo, ``Facehop: A light-weight low-resolution face gender classification method,'' in \emph{International Conference on Pattern Recognition}.\hskip 1em plus 0.5em minus 0.4em\relax Springer, 2021, pp. 169--183.

\bibitem{chen2021defakehop}
H.-S. Chen, M.~Rouhsedaghat, H.~Ghani, S.~Hu, S.~You, and C.-C.~J. Kuo, ``Defakehop: A light-weight high-performance deepfake detector,'' in \emph{2021 IEEE International Conference on Multimedia and Expo (ICME)}.\hskip 1em plus 0.5em minus 0.4em\relax IEEE, 2021, pp. 1--6.

\bibitem{chen2022defakehop++}
H.-S. Chen, S.~Hu, S.~You, C.-C.~J. Kuo \emph{et~al.}, ``Defakehop++: An enhanced lightweight deepfake detector,'' \emph{APSIPA Transactions on Signal and Information Processing}, vol.~11, no.~2, 2022.

\bibitem{mei2022greenbiqa}
Z.~Mei, Y.-C. Wang, X.~He, and C.-C.~J. Kuo, ``Greenbiqa: A lightweight blind image quality assessment method,'' in \emph{2022 IEEE 24th International Workshop on Multimedia Signal Processing (MMSP)}.\hskip 1em plus 0.5em minus 0.4em\relax IEEE, 2022, pp. 1--6.

\bibitem{zhang2021anomalyhop}
K.~Zhang, B.~Wang, W.~Wang, F.~Sohrab, M.~Gabbouj, and C.-C.~J. Kuo, ``Anomalyhop: an ssl-based image anomaly localization method,'' in \emph{2021 International Conference on Visual Communications and Image Processing (VCIP)}.\hskip 1em plus 0.5em minus 0.4em\relax IEEE, 2021, pp. 1--5.

\bibitem{zhu2022pixelhop}
Y.~Zhu, X.~Wang, H.-S. Chen, R.~Salloum, and C.-C.~J. Kuo, ``A-pixelhop: A green, robust and explainable fake-image detector,'' in \emph{ICASSP 2022-2022 IEEE International Conference on Acoustics, Speech and Signal Processing (ICASSP)}.\hskip 1em plus 0.5em minus 0.4em\relax IEEE, 2022, pp. 8947--8951.

\bibitem{RGGID}
Y.~Zhu, X.~Wang, R.~Salloum, H.-S. Chen, and C.-C. Kuo, ``Rggid: A robust and green gan-fake image detector,'' \emph{APSIPA Transactions on Signal and Information Processing}, vol.~11, 01 2022.

\bibitem{zhu2023green}
Y.~Zhu, X.~Wang, H.-S. Chen, R.~Salloum, and C.-C.~J. Kuo, ``Green steganalyzer: A green learning approach to image steganalysis,'' \emph{arXiv e-prints}, pp. arXiv--2306, 2023.

\bibitem{xie2022graphhop}
T.~Xie, B.~Wang, and C.-C.~J. Kuo, ``Graphhop: An enhanced label propagation method for node classification,'' \emph{IEEE Transactions on Neural Networks and Learning Systems}, 2022.

\bibitem{xie2023label}
T.~Xie, R.~Kannan, and C.-C.~J. Kuo, ``Label efficient regularization and propagation for graph node classification,'' \emph{IEEE Transactions on Pattern Analysis and Machine Intelligence}, 2023.

\bibitem{liu2021voxelhop}
X.~Liu, F.~Xing, C.~Yang, C.-C.~J. Kuo, S.~Babu, G.~El~Fakhri, T.~Jenkins, and J.~Woo, ``Voxelhop: Successive subspace learning for als disease classification using structural mri,'' \emph{IEEE journal of biomedical and health informatics}, vol.~26, no.~3, pp. 1128--1139, 2021.

\bibitem{zhang2020pointhop}
M.~Zhang, H.~You, P.~Kadam, S.~Liu, and C.-C.~J. Kuo, ``Pointhop: An explainable machine learning method for point cloud classification,'' \emph{IEEE Transactions on Multimedia}, 2020.

\bibitem{kadam2022r}
P.~Kadam, M.~Zhang, S.~Liu, and C.-C.~J. Kuo, ``R-pointhop: A green, accurate, and unsupervised point cloud registration method,'' \emph{IEEE Transactions on Image Processing}, vol.~31, pp. 2710--2725, 2022.

\bibitem{zhang2022gsip}
M.~Zhang, P.~Kadam, S.~Liu, and C.-C.~J. Kuo, ``Gsip: Green semantic segmentation of large-scale indoor point clouds,'' \emph{Pattern Recognition Letters}, vol. 164, pp. 9--15, 2022.

\bibitem{kadam2022greenpco}
P.~Kadam, M.~Zhang, J.~Gu, S.~Liu, and C.-C.~J. Kuo, ``{GreenPCO}: An unsupervised lightweight point cloud odometry method,'' in \emph{2022 IEEE 24th International Workshop on Multimedia Signal Processing (MMSP)}.\hskip 1em plus 0.5em minus 0.4em\relax IEEE, 2022, pp. 01--06.

\bibitem{kadam2023s3i}
P.~Kadam, H.~Prajapati, M.~Zhang, J.~Xue, S.~Liu, and C.-C.~J. Kuo, ``{S3I-PointHop}: {SO(3)}-invariant pointhop for 3d point cloud classification,'' in \emph{ICASSP 2023-2023 IEEE International Conference on Acoustics, Speech and Signal Processing (ICASSP)}.\hskip 1em plus 0.5em minus 0.4em\relax IEEE, 2023, pp. 1--5.

\bibitem{xu2017understanding}
H.~Xu, Y.~Chen, R.~Lin, and C.-C.~J. Kuo, ``Understanding cnn via deep features analysis,'' in \emph{2017 Asia-Pacific Signal and Information Processing Association Annual Summit and Conference (APSIPA ASC)}.\hskip 1em plus 0.5em minus 0.4em\relax IEEE, 2017, pp. 1052--1060.

\bibitem{HOG}
N.~Dalal and B.~Triggs, ``Histograms of oriented gradients for human detection,'' in \emph{2005 IEEE Computer Society Conference on Computer Vision and Pattern Recognition (CVPR'05)}, vol.~1, 2005, pp. 886--893 vol. 1.

\bibitem{yang2022design}
Y.~Yang, H.~Fu, and C.-C.~J. Kuo, ``Design of supervision-scalable learning systems: Methodology and performance benchmarking,'' \emph{arXiv preprint arXiv:2206.09061}, 2022.

\bibitem{MNIST}
Y.~LeCun, L.~Bottou, Y.~Bengio, and P.~Haffner, ``Gradient-based learning applied to document recognition,'' \emph{Proceedings of the IEEE}, vol.~86, no.~11, pp. 2278--2324, 1998.

\bibitem{xiao2017fashion}
H.~Xiao, K.~Rasul, and R.~Vollgraf, ``Fashion-mnist: a novel image dataset for benchmarking machine learning algorithms,'' \emph{arXiv preprint arXiv:1708.07747}, 2017.

\bibitem{cifar10}
A.~Krizhevsky and G.~Hinton, ``Learning multiple layers of features from tiny images,'' University of Toronto, Toronto, Ontario, Tech. Rep., 2009.

\end{thebibliography}

\end{document}